\documentclass[twocolumn]{revtex4}
\usepackage{graphicx,amsmath}

\catcode`\é=13\def é{\'e} \catcode`\è=13\def è{\`e}
\catcode`\ù=13\def ù{\`u} \catcode`\à=13\def à{\`a}
\catcode`\ê=13\def ê{\^e} \catcode`\â=13\def â{\^a}
\catcode`\ô=13\def ô{\^o} \catcode`\û=13\def û{\^u}
\catcode`\î=13\def î{\^\i} \catcode`\ç=13\def ç{\c{c}}
\catcode`\ï=13\def ï{\"{i}}

\newcommand{\be}{\begin{equation}}
\newcommand{\ee}{\end{equation}}
\newcommand{\bea}{\begin{eqnarray}}
\newcommand{\eea}{\end{eqnarray}}

\begin{document}

\title{Tuning the proximity effect in a superconductor-graphene-superconductor junction}
\author{C. M. Ojeda-Aristizàbal, M. Ferrier, S. Guéron and H. Bouchiat}
\affiliation{Laboratoire de Physique des Solides,
Univ. Paris-Sud, CNRS, UMR 8502, F-91405, Orsay, France}
\begin{abstract}
We have tuned in situ the proximity effect in a single graphene layer  coupled to two Pt/Ta superconducting electrodes. An annealing current through the device changed the transmission coefficient of the electrode/graphene interface, increasing the probability of multiple Andreev reflections. Repeated annealing steps improved the contact sufficiently for a Josephson current to be induced in graphene. 
\end{abstract}

\today
\maketitle

\section{Introduction}
Graphene, the one atom thick crystal of carbon atoms, is a unique material due to its electronic band structure, in which the charge carriers behave as massless particles. Electrostatic gating can tune the density of carriers, and change the sign of their charge when the electron-hole symmetry point (the so-called Dirac point) is crossed \cite{Novoselov}. The consequences of graphene's unique band structure are many, in particular an unconventional Quantum Hall effect, which was discovered in the very first measurements on graphene \cite{GeimHall}. Another consequence is a special type of Andreev reflection at the interface between graphene and a superconductor, when the Fermi energy lies within $\Delta$, the superconducting gap, of the Dirac point. Whereas in conventional Andreev reflection, both the electron and hole  of the Andreev pair belong to the conduction band, in graphene close to the Dirac point, an electron of the conduction band can be reflected into a hole from the valence band , leading to specular reflection instead of the usual retroreflection \cite{Beenakker}. Experiments on graphene connected to superconducting electrodes \cite{AndreiSuperI,Heersche} have so far not managed to observe this original Andreev reflection. One of the reasons is that spatial inhomogeneities in Fermi energy are larger than $\Delta$, given the superconductors used (mostly Al). The observation of the special Andreev reflection thus requires a combination of superconducting electrodes with larger gaps and lower local doping, thus cleaner graphene samples. One way of improving the quality of the graphene samples is current annealing. It was shown that for graphene on a substrate, annealing displaces of the Dirac point to low voltage and increase the sample homogeneity. This was linked to the migration of adsorbed impurities to the edges of the graphene sheet \cite{Bachtold}. In this paper, we investigate the effect of current annealing on the proximity effect in  a graphene sheet connected to tantalum, a superconductor different from previous experiments. We find that annealing increases the graphene/superconductor contact, and gradually changes the proximity effect from one with a low bias peak of resistance to one in which a supercurrent is induced in the graphene. Annealing also changes the visibility of multiple Andreev reflexion (MAR) peaks. We also report on the effect of the superconducting electrodes on the conductance fluctuations of the graphene sheet.

\section{Sample fabrication}
The superconductor-graphene-superconductor junction (SGS) was fabricated with exfoliated graphene deposited on a doped silicon substrate with a 285 nm thick oxide, which allows the visual detection with an optical microscope, while providing a capacitively coupled gate electrode.
Raman spectroscopy confirmed that the sample was made of a single layer graphene. The leads,  a Pt/Ta/Pt trilayer of thicknesses 3 nm/70 nm/3 nm, were made using standard electron beam lithography and lift-off. Platinum and tantalum were sputter deposited. The distance between the electrodes L (figure \ref{UCF}) is  about 330 nm, and the width of the junction W is $2.7\mu m$. The critical temperature of the Ta leads is 2.5 K, and the critical field is 2 Tesla.

\begin{figure}[h]
    \includegraphics[width=0.5\textwidth]{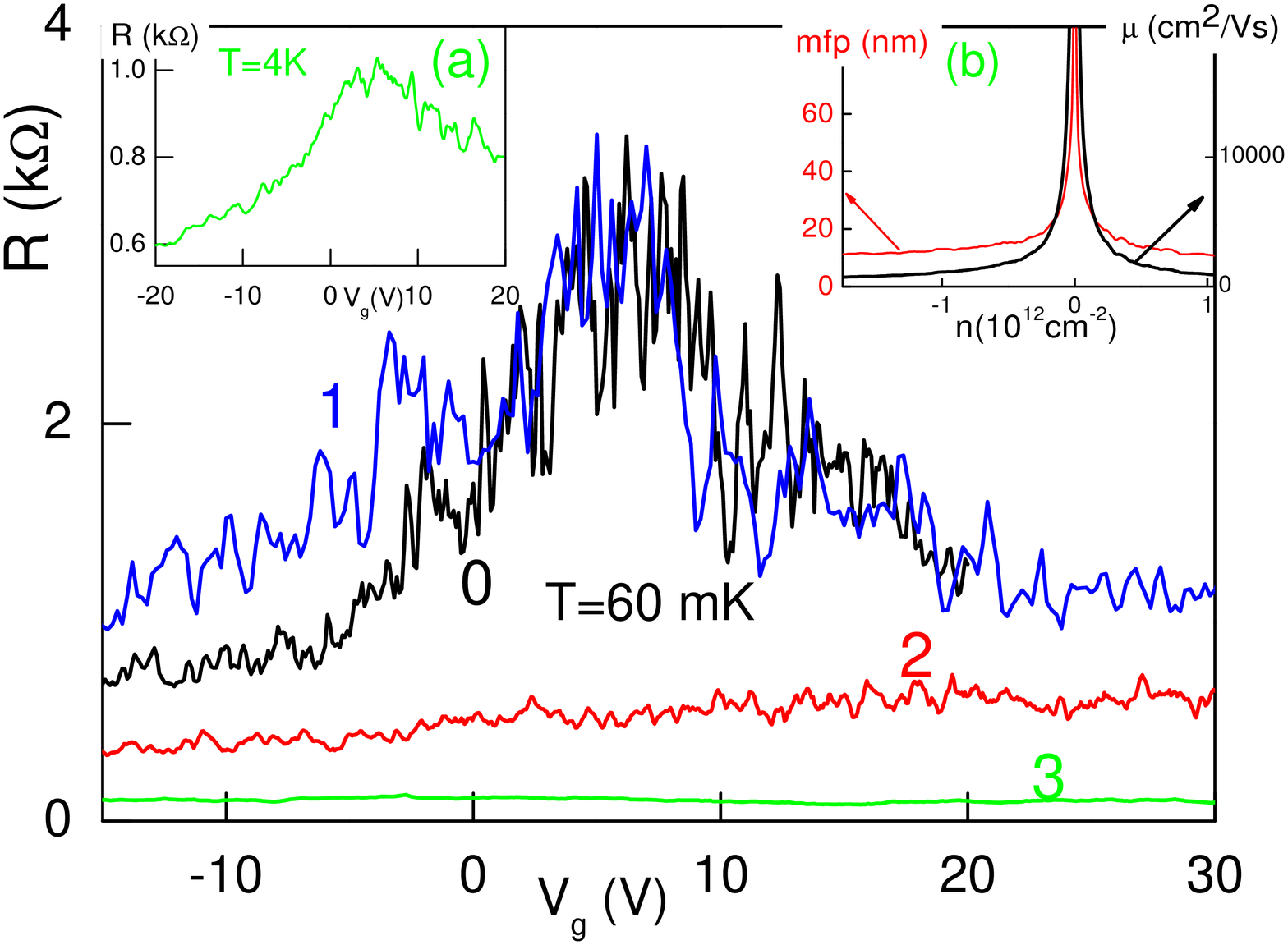}
    \caption{Gate voltage dependence of the two-wire resistance of the sample before and after different annealing steps, at 60 mK. Annealing 1, 2, and 3 were implemented with 3, 6, and 10 mA for several minutes. The temperature of the dilution refrigerator varied from 60 mK to 10 K during annealing. The last annealing step (curve 3) induced a full proximity effect in the sample: a supercurrent ran through the graphene. The gate dependence 3 shown here corresponds to this last state, but with a 200 G magnetic field applied which destroys the proximity effect and thus measures the intrinsic sample resistance in this final stage.
    The "noisy'' resistance curves are actually reproducible conductance fluctuations (see text). Inset (a): Resistance versus gate voltage before any annealing, at 4.2 K; Inset (b): mobility and mean free path of the graphene sheet before annealing, deduced from the 4.2 K curve (inset a).   }
    \label{Fig1}
\end{figure}
 
\section{Transport Measurements and current annealing steps}  
The measurements were performed in a dilution refrigerator with a base temperature of 60 mK, via lines with room temperature low pass filters. Two terminal differential resistance measurements were implemented with a lock-in amplifier, applying a small ac current (50nA) superimposed on a dc current. The carrier density was controlled by applying a voltage to the doped silicon back gate. 

Figure \ref{Fig1} shows the gate voltage dependence of the sample resistance, before annealing (curve 0) and after three annealing steps (curves 1 to 3).  In the first annealing step we applied a 3 mA current through the sample for three minutes, which corresponds to a current density of $2*10^{8}A/cm^{2}$ if we take the graphene thickness to be 0.36 nm. The second and third annealing steps were implemented with 6 mA and 10 mA respectively. 
As  seen in figure \ref{Fig1}, in the first two curves (curves 0 and 1), the Dirac point, voltage region in which the resistance is maximum because the carrier density is minimal, is located at 5 V. This slight offset is attributed to doping by charged impurities on the graphene or between the graphene and the substrate. After the second annealing step (curve 2), we find that the Dirac point has shifted to about 20 V and that the resistance has decreased by more than a factor two everywhere, and by up to a factor six around the original Dirac point.
Finally the last annealing step decreased the resistance yet further (curve 3), and led to a full proximity effect, with a zero resistance of the sample at low enough current bias. 
Since a two wire resistance is the sum of the intrinsic resistance of the graphene sheet and the contact resistance between the graphene and the metal electrodes, one cannot from one curve alone deduce the relative contribution of each. However the qualitative difference between the second and third annealing steps (curves 2 and 3), in which a full proximity effect is induced in the graphene, indicates that annealing must have greatly improved the quality of the graphene/contact interface, in addition to increasing the mean free path and changing the doping. Indeed, an increased doping and larger mean free path alone would not cause the appearance of a supercurrent, but would merely increase the value of an already existing critical current. Only an improved interface transparency could change qualitatively a proximity effect from one without supercurrent to one with a supercurrent. 
Inset (b) of figure \ref{Fig1} presents the  mean free path $l_e=h\sigma/(2k_{F}e^{2})$ and mobility $\mu=\sigma\pi/ek_{F}^{2}$ of the sample deduced using a plane capacitor model, and extracted from the conductance versus gate voltage curve at 4.2 K, before annealing (inset (a)). Here the Fermi wavevector is $k_{F}=\sqrt{\epsilon_{r}\epsilon_{0}\pi/ed}\sqrt{V_{g}-V_{Dirac}}$. If we do not include an interface resistance, we find a mean free path of roughly $l_e=15$ nm, corresponding to diffusive transport.  We also find a mobility of about 2000 $cm^{2}/Vs$ away from the Dirac point, at a density of $5.10^{11}~cm^{-2}$, which is lower than found by other groups (roughly 20 000 $cm^{2}/Vs$ \cite{AndreiSuperI}). This difference in mobilities can be partially attributed to the contact resistance which lowers the apparent mean free path and mobility estimated from the total resistance. The diffusive nature of transport, as for all samples on substrate, is attributed to scattering from impurities in graphene and defects between the $SiO_{2}$ substrate and the graphene.
The spatial inhomogeneity in doping is great, as seen in the gate-voltage width of the Dirac point, which translates into an inhomogeneity of the Fermi energy of about 85 meV, using conversion of gate voltage into Fermi energy via the plane capacitor model,  $E=30\sqrt{V_{g}}$ meV. 

\begin{figure}[h]
    \includegraphics[width=0.5\textwidth]{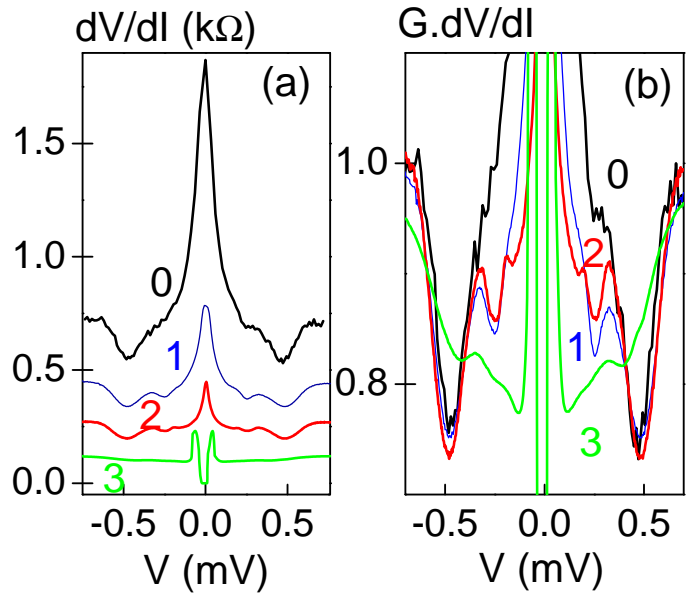}
    \caption{(a) Differential resistance versus bias voltage as a function of annealing steps. The curves are taken at gate voltages of 0,-20, -16 and 15 V respectively for curves 0 to 3, and are not shifted vertically. In curve 3 the 40 $\Omega$ resistance of the wires leading to the sample has been subtracted.  (b) Differential resistance normalized to the 0.75 meV value, zoom around the MAR region. A supercurrent appears after the third annealing step (curve 3). The MAR structures demonstrate the fact that the interface transparency is not perfect, and that it improves with annealing.
    }
    \label{evoldvdi}
\end{figure}     

We now turn to the effect of annealing on the proximity effect induced in graphene. 
As the differential resistance curves of figure \ref{evoldvdi} show, the effect of annealing is to reduce the S/graphene/S junction resistance over the entire bias voltage range. In particular, the zero bias resistance decreases strongly, and goes from a peak to a dip: after annealing 3 a full proximity effect is induced in the sample, as seen from the zero resistance state at zero voltage in curve 3. Panel b plots the differential resistance curves normalized by their high bias value to emphasize the resistance dips at voltages of 170, 260, 480 $\mu$V.  These values are close to $2e\Delta /n$, with n= 1,2,3 and $2\Delta=500~\mu$V. A fourth dip at smaller voltage of 90 $\mu$V is in between n=5 and 6. The principle dips can be attributed to multiple Andreev reflexions (MAR) occurring at the graphene/superconducting electrode interfaces. The higher order MAR peaks become clearer as the interface transparency improves (curves 0, 1 and 2), as expected since an increased transparency enables higher order tunneling processes. The MAR dips are still visible in the proximity induced superconducting state, although they are smeared out, as expected for an SNS junction with a high transparency, see \cite{Flensberg}. Since a perfect interface leads to a supercurrent and no subgap structure, the small residual subgap structure after annealing 3, when a supercurrent is induced, is a proof of a still imperfect interface transparency. An additional reason for the increase in MAR visibility after annealings 1 and 2 is the decrease (in width and amplitude) of the central (low bias) resistance peak. A quantitative comparison, which could yield the exact transparency at each stage, would require the adaptation of the OBTK theory (\cite{OBTK,Flensberg}) to the case of diffusive SNS junctions with a finite interface transparency, or better yet including the specificities of graphene. To our knowledge such calculations do not exist yet \cite{Hammer,Cuevasgraphene}.

\begin{figure}[h]
\includegraphics[width=3in]{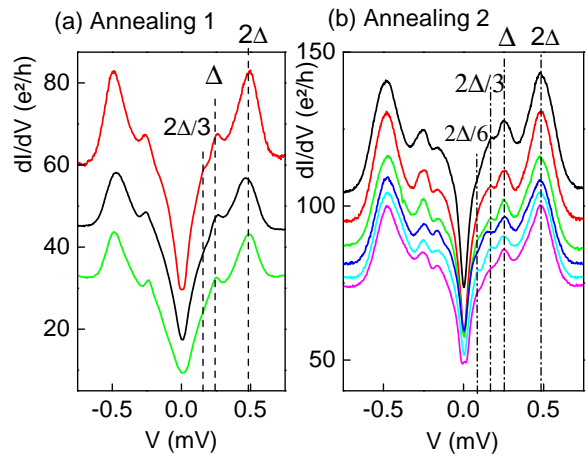}  
\caption{MAR seen in the differential conductance as a function of bias voltage after annealing 1 and 2, at 60 mK. The gate voltage does not affect the curves qualitatively ($V_g=-3, -8, 5$ V for panel (a), and $V_g=-24,-22, -18, -16, -15$ and -6 V in panel (b)). Up to four MAR peaks are seen after the second annealing step. Note that curves have not been shifted vertically}
\label{Cuiit34}
\end{figure}

Figure \ref{Cuiit34} shows that the gate voltage does not change qualitatively the differential conductance curves, and shows that the number of visible MAR increases with annealing. We interpret this as due to the larger contact transparency.

The value of the superconducting gap $\Delta =250~\mu eV$ deduced from the position of the MAR resistance dips is the same as measured in a tunnel junction formed between a different graphene sheet and a similar Pt/Ta/Pt trilayer (different experiment, not shown). This value is smaller than the gap extracted from the critical temperature measured, using the BCS formula $\Delta_{BCS}=1.76*k_{B}T_{c}=379$ $\mu$V, with $T_{c}$=2.5 K. This can be attributed to the 3 nm-thick platinum layer deposited between graphene and the thick tantalum layer. Indeed, it is known that the measured $T_{c}$ of a Pt/Ta bilayer with thick Ta is practically the bulk $T_c$ of Ta whereas the gap at the bottom of the Pt layer may be much smaller than the bulk $T_c$ of Ta \cite{Takis}.

\begin{figure}[h]
\includegraphics[width=3.3 in]{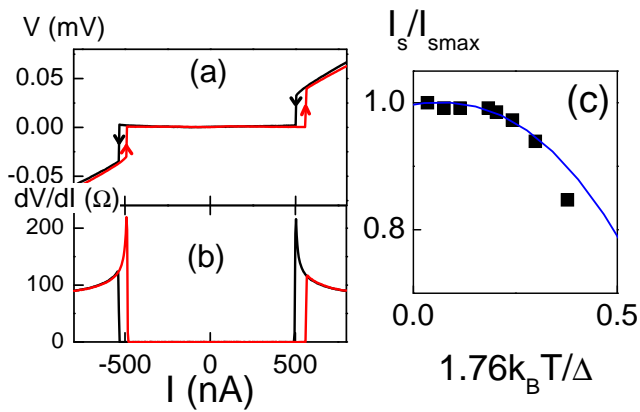}
\caption{Full proximity effect induced in graphene after the third annealing step. (a) I-V curve and (b)dV/dI(I) of the SGS junction taken at 60 mK: a zero resistance state prevails at bias currents below a switching current of 600 nA. (c) The temperature dependance of the switching current (data points) follows a Kulik-Omelyanchuk law (continuous line) typical of short SNS junctions, see text. We have used the gap $\Delta=250~\mu V$ deduced from the MAR features.}
\label{SuperI}
\end{figure}

We now turn to the supercurrent induced in graphene by the third annealing step.
Figure \ref{SuperI} shows the I(V) curve at 60 mK and $V_g=15.5$ V, with a zero resistance state for currents smaller than the switching current $I_s=600$ nA, and a linear I(V) curve above. The corresponding normal resistance is $R_N=90$ $\Omega$. The switching current varies from 720 nA at $V_g=-64$ V to 480 nA at $V_g=64$ V, and $R_N$ varies from 80 to 105 $\Omega$. The product $R_NI_s$ thus varies between 58 and 50 $\mu$V, which is roughly $\Delta /5e$. The predictions for the value of $R_NI_s$ in any SNS junction differ depending on whether the junction is in a short or long junction limit, i.e. whether the junction length L is much smaller or much greater than the superconducting coherence length $\xi =\sqrt{\hbar D/\Delta}$. Here $D$ is the diffusion constant $D=v_{F}l_e/2$ and $l_e$ is the elastic mean free path in graphene after annealing 3. At that stage the Dirac point is not clearly defined, but roughly corresponds to a gate voltage of 65 V. The mean free path deduced at a gate voltage of 15.5 V is then $l_e=55~nm$, which yields $\xi=260~nm$, of the order of the distance between contacts L. Thus the sample is in the intermediate regime between short and long junction, and the Thouless energy $E_{Th}=\hbar D/L^2$ is of the same order of magnitude as the superconducting gap. The temperature dependence of the switching current also points to a rather short junction limit, since it follows a Kulik Omelyanchuck-like dependence \cite{Likharev}(see fig \ref{SuperI} (c)).

The ratio $L/\xi=1.3$ leads to a theoretical $R_NI_s$ product of 1.3 $\Delta /e$ for a perfect interface \cite{Wilhem}, a factor six higher than what is measured (other experiments also find less than expected, by roughly a factor two \cite{Heersche,AndreiSuperI}). The discrepancy is too large to be explained solely by an interface resistance \cite{Hammer}, since a factor of 6 reduction of $R_NI_s$ with respect to the expected value corresponds in short junctions to an interface resistance many times the graphene resistance. In addition to the interface resistance, dephasing by fluctuators on and beneath the graphene, as well as the electromagnetic environment may cause the smaller than expected measured switching current.

\begin{figure}[h]
    \includegraphics[width=3 in]{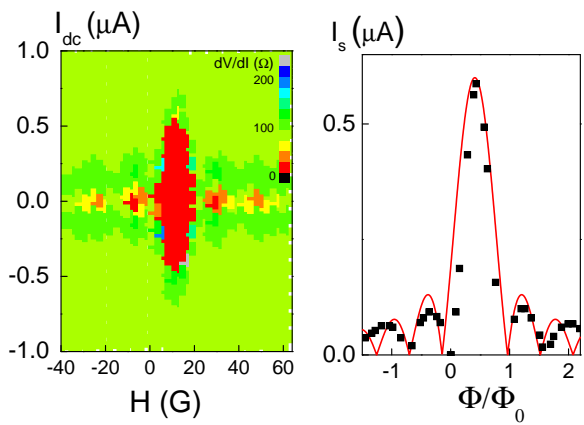}
    \caption{Switching current in a magnetic field perpendicular to the the graphene plane. Left panel: color coded differential resistance as a function of bias current and applied magnetic field. Right panel: Field dependence of the switching current, extracted from the differential resistance curves (Data points), compared to a Fraunhofer diffraction pattern. The calculated Fraunhofer pattern has been adjusted to take into account a residual magnetic field of 13 G, and an effective area larger by a factor 1.8 than the graphene area between the electrodes (see text).}
    \label{Fraunhofer}
\end{figure}

A question that naturally arises is whether the induced supercurrent could be caused by the diffusion onto the graphene sheet of superconducting grains during the annealing process, since the large temperatures reached may increase the mobility of atoms tremendously. These atoms could then form a superconducting weak link, through which a supercurrent would flow. The experimental answer to this question is given by the field dependence of the switching current, shown in Fig. \ref{Fraunhofer}.

The figure shows that the switching current is modulated by the magnetic field (applied perpendicularly to the graphene plane) according to an interference pattern that resembles the Fraunhofer pattern found in rectangular superconductor-normal metal-superconductor junctions \cite{Cuevas}. The fit is not perfect, in particular the effective sample area must be increased by a factor two to fit the experimental data. This larger effective area may be explained by a finite penetration depth, and non local trajectories in the graphene sheet beyond the superconducting electrodes (see sample picture in Fig \ref{Fraunhofer}). The penetration depth in a perpendicular magnetic field in a disordered superconductor is given by $\Lambda_\perp=\Lambda_0^2\xi_0/(l_ed)$, where $\Lambda_0=\sqrt{m/(ne^2\mu_0)}$ is the London penetration depth in a clean metal, $\xi_0=\hbar v_F/(\pi \Delta)$ is the clean superconducting coherence length, $l_e$ the mean free path in the superconductor, $d$ the superconductor thickness, and $n$ the electron density \cite{Tinkham}. This yields a perpendicular penetration depth of 120 nm for our sample, and including this length on each superconducting electrode practically doubles the effective normal surface. In conclusion, the field periodicity of the interference pattern excludes the possibility of a superconducting Ta weak link crossing the graphene.

\section{Conductance fluctuations}
Universal conductance fluctuations are typical of phase coherent samples, and have been investigated mostly in metals and two dimensional electron gases made of semiconducting heterostructures. The conductance of a sample can fluctuate as a function of magnetic field, bias voltage, and gate voltage. The amplitude of fluctuations in the normal state depends on the dimensionality of the sample \cite{Skocpol}. For a wire shorter than the phase coherence length $L_\varphi$, the fluctuation amplitude is universal and of the order of $e^2/h$. In a two dimensional sample of width W and length L, the fluctuation amplitude has been shown to be given by $\sqrt{(max(L_\varphi,W)/L)}(L_\varphi/L)e^2/h$ which in the case of a sample of width greater than $L_\varphi$ yields $\sqrt{W/L} (L_\varphi/L)e^2/h$. 

Predictions differ about the exact ratio between fluctuations in a NS system and the same system in the normal state \cite{BeenakkerRMP,Lambert}. The prediction by Beenakker et al. are that the fluctuations in a NS system should be twice those in the NN system, in zero field, $\delta G_{NS}/\delta G_{NN}=2$, and $2\sqrt{2}$ in a magnetic field greater than the coherence field $B_c=\Phi_0/L_{\varphi} ^2$. Such predictions were checked experimentally in a semiconducting nanowire \cite{deFranceschi}. 

The case of graphene has just recently come into consideration, and numerical simulations suggest that these UCF should not be universal in graphene, because of the different nature of scattering induced by impurities \cite{BeenakkerUCFgraphene}.

\begin{figure}[h]
    \includegraphics[width=0.45\textwidth]{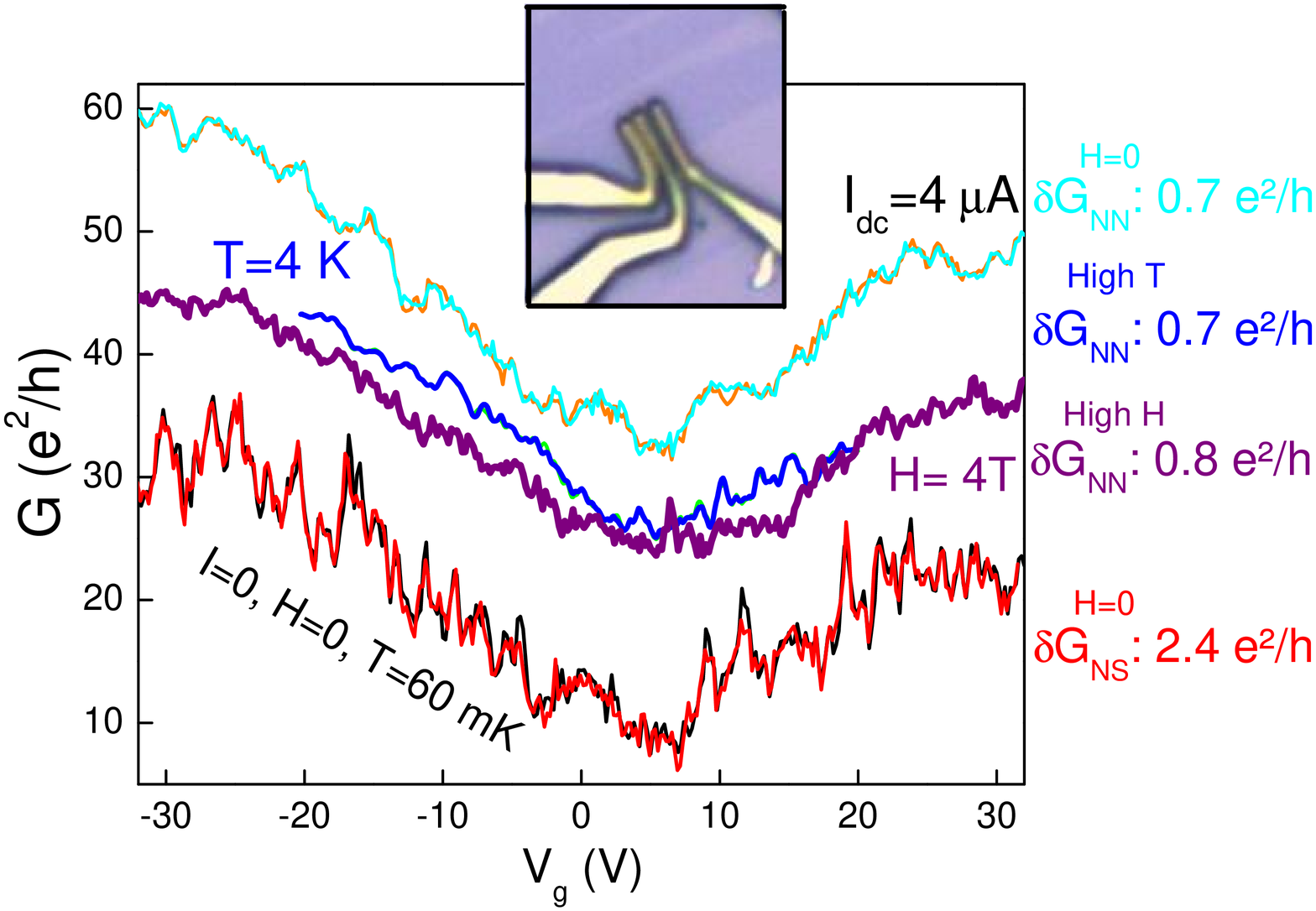}
    \caption{Gate voltage dependence of conductance after annealing 1, in different conditions of temperature, bias current, and magnetic field. The fluctuations are reproducible, as shown by the fact that both increasing and decreasing gate voltage sweep yields the same fluctuations. The standard deviation, deduced from the high pass filtered curve, is $2.4~e^2/h$ for the low temperature zero field curve, in which the electrodes are superconducting;  $0.8~e^2/h$ for the low temperature curve at high field (4 T), above the critical field of the superconductor,  and $0.7~e^2/h$ for the low temperature zero field curve with a current bias above the critical current of the electrode. The fluctuations are $0.7 e^2/h$ for the curve at 4.2 K.
Inset: Optical image of the sample.}
    \label{UCF}
\end{figure}

The conductance fluctuations after annealing 1 as a function of gate voltage are plotted in Fig. \ref{UCF}, at low temperature and zero field (case of a coherent NS system at low temperature), 4 T (coherent NN system at low temperature and high field), and with a dc current applied to the sample (NN system in zero field), and also at 4.2 K (NN system at high temperature, shorter coherence length). The extracted standard deviation is $2.4~e^2/h$ for the low temperature zero field curve, in which the electrodes are superconducting; it is $0.8~e^2/h$ for the low temperature curve at high field (4T), above the critical field of the superconductor,  and $0.7~e^2/h$ for the low temperature zero field curve with a current bias above the critical current of the electrode. The fluctuations are $0.7~e^2/h$ for the curve at 4.2 K. In comparison, the conductance fluctuations of a phase coherent NN sample with the aspect ratio of the present experiment should be $\delta G_{NN}=3~e^2/h$. We thus find fluctuation which are smaller than that value. But we find a factor of three enhancement of the fluctuations with the electrodes in their superconducting state compared to  when the electrodes are in the normal state, in good agreement with the theoretical prediction. A quantitative comparison requires a better characterized interface transparency, and the evaluation of the phase coherence length in the sample after the first annealing procedure. 

Finally, an interesting feature of the gate voltage dependence of the fluctuations is their typical energy scale of 1 V. This corresponds to a typical variation of Fermi energy of 15 meV, which translates in a typical length scale of 50 nm. By analogy with the universal conductance fluctuations whose typical energy corresponds to the phase coherence length, we conjecture that this second, smaller length scale, which appears in the reproducible fluctuations in graphene, corresponds to the typical size of the so-called puddles of graphene. Such electron and hole-doped regions have been visualized in near probe spectroscopy \cite{Martin, Deshpande}, but have not yet to our knowledge been inferred from their mesoscopic signature. This question will be described in details elsewhere \cite{Monteverde}.

\section{Conclusion}
In conclusion, we have tuned the proximity effect in a graphene sheet by running a large current through the sample. The annealing improved the quality of the graphene/electrode interface, and changed the resistance from a low bias peak to a zero-resistance superconducting state. The Dirac point was not sufficiently well defined in that state to check the predicted original properties of the proximity effect in S-graphene-S junctions \cite{Titov}. A promising possibility would be to perform this kind of annealing on suspended sample, to improve the sample mobility and the homogeneity of doping \cite{Bachtold}. 

\section{Acknowledgments}
We thank M. Cazayous, Y. Gallais and A. Sacuto for help with the Raman spectroscopy measurements, and T. Kontos, M. Aprili, and J. C. Cuevas for discussions. C. O. is funded by CE program CEE MEST CT2004 514307 EMERGENT CONDMAT PHYS Orsay, and this research was supported by the European program HYSWICH.

\end{document}